\documentstyle[aps,twocolumn,epsfig]{revtex}

\makeatletter

\def\subsection{\@mainheadfalse
\@startsection{subsection}{2}{\z@}{0.8cm plus1ex minus
 .2ex}{0.5cm plus1ex minus.2ex}{\noindent\reset@font\small\it}}
\makeatother

\begin{document}
  \title{{\tiny JOURNAL OF COMPUTATIONAL PHYSICS}
    \hfill {\sl Submitted}\\~~\\~~\\~~\\
    {\LARGE\bf\sf Adaptive Mesh Refinement for Singular Current Sheets} \\
    {\LARGE\bf\sf in Incompressible Magnetohydrodynamic Flows}}

  \author{\sc Holger Friedel, Rainer Grauer, and Christiane Marliani}

  \address{~~\\Institut f{\"u}r Theoretische Physik I \\
           Heinrich-Heine-Universit{\"a}t D{\"u}sseldorf \\
           D-40225 D{\"u}sseldorf, Germany}
 \maketitle
 \narrowtext

\noindent\hrulefill

\medskip
{\sf
 The formation of current sheets in ideal incompressible
 magnetohydrodynamic flows in two dimensions is studied numerically
 using the technique of adaptive mesh refinement. The growth of
 current density is in agreement with simple scaling assumptions. As
 expected, adaptive mesh refinement shows to be very efficient for
 studying singular structures compared to non-adaptive treatments.
}

\noindent\hrulefill
\section{Introduction}

The formation of singularities in hydro- and magnetohydrodynamic flows
is still a controversial issue in the mathematics and physics
community. Since mathematically only very little is known
\cite{beale-kato-etal:1984}, one has to rely on numerical simulations.
Even in very elaborate numerical experiments (see Bell and
Marcus~\cite{bell-marcus:1992b}, Kerr~\cite{kerr:1993}) non-adaptive
treatment is limited very soon by the computer memory available,
resulting in a resolution of less than 512 grid points in each spatial
direction.  Since the singular structures like tubes and sheets are
not space filling, adaptive mesh codes seem to be the right choice for
studying these problems, as has been done by Pumir and
Siggia\cite{pumir-siggia:1990,pumir-siggia:1992}. Unfortunately, the
methods used in \cite{pumir-siggia:1990,pumir-siggia:1992} could only
refine the region around a singular point which lead in
\cite{pumir-siggia:1990} to a substantial loss of energy.  Of course,
it is desired to refine all regions where the numerical resolution is
insufficient. Modern adaptive mesh refinement algorithms, as
introduced by Berger and Colella~\cite{berger-colella:1989} and Bell
{\it et al.}~\cite{bell-berger-etal:1994}, do not possess the above
limitations and are good candidates for studying singularity formation
even in incompressible systems.

In this paper, we investigate the formation of singular current sheets
described by the ideal incompressible magnetohydrodynamic equations
(MHD equations) in two dimensions for the time evolution of the
velocity field ${\bf u}$ and magnetic field ${\bf B}$. Using
Els\"asser variables ${\bf z}^\pm = {\bf u} \pm {\bf B}$, the
MHD equations take the symmetric form
\begin{equation}
  \partial_t {\bf z}^\pm + {\bf z}^\mp \cdot \nabla {\bf z}^\pm 
  + \nabla p = 0  \; , \;\; \mbox{div} \; {\bf z}^\pm = 0 \;\; .
  \label{mhdz} 
\end{equation}
The equations (\ref{mhdz}) are integrated in a periodic quadratic box
of length $2 \pi$ using adaptive mesh refinement with rectangular
grids self-adjusting to the flow. In each rectangular grid, a
projection method is used where the time-stepping is performed in a
second order upwind
manner~\cite{bell-colella-etal:1989,bell-marcus:1992,grauer-marliani:1995}.
For the projection step, we need the vorticities $\omega^\pm = (\nabla
\times{\bf z}^\pm) \cdot {\bf e}_z$ and potentials $\psi^\pm$ which
are related by $\Delta \psi^\pm = \omega^\pm$.

The outline of the paper is as follows. In the next section, the
adaptive mesh refinement algorithm is introduced. Then, we discuss the
numerical results and compare the growth of current density with the
prediction of Sulem {\it et al.}~\cite{sulem-frisch-etal:1985}.
Finally, we conclude that adaptive mesh refinement is an ideal tool
for studying singular structures and should be pursued further to
study three dimensional problems as the finite time blow up in the
incompressible Euler equations.

\section{Adaptive Mesh Refinement}
\subsection{General Strategy}

The main idea of adaptive mesh refinement is simple. One starts with a
grid of given resolution and integrates the partial differential
equation as usual. As soon as some criterion is fulfilled, this
initial grid is refined. This is done by marking all critical grid
points where the discretization error exceeds a prescribed value. Then
new grids with finer resolution and timestep are generated which cover
all these critical points. These grids belonging to the next level are
then filled with interpolated data from the first level.  Then, one
integrates both levels until the resolution again becomes
insufficient. Now the critical points are collected over all grids of
the actual level being refined. Filling the new grids with data is
achieved by first taking data from the previous level and, if
existing, data from former grids of the same resolution. This process
is repeated recursively. In addition, to communicate the boundary
conditions, each grid needs information about its parent grids and its
neighbors. As one can see already, adaptive mesh refinement requires
the management of lists of levels, critical points, grids, parent
grids and neighbors. Therefore, we programmed the handling of those
structures in C++ whereas the numerically expensive integrations are
done in fortran. In order to encourage the reader to use adaptive mesh
refinement we describe the above outline in more detail in the next
paragraphs.

To deal with all the different lists we defined templated list classes
and iterators which can be used for all classes representing levels,
grids, critical points, parents and neighbors.

The integrator used for all grids is based on a projection method
combined with second order upwinding. This scheme motivated by Bell
{\it et al.}~\cite{bell-colella-etal:1989} was previously applied to
incompressible magnetohydrodynamic flows in two
dimensions~\cite{grauer-marliani:1995}. It is clear, that the
equations under consideration can be easily exchanged by other ones
using an explicit algorithm since the structures needed for adaptive
mesh refinement and the integrator are independent of each other.

The timestep on a given {\em level} is advanced as illustrated by the
following piece of pseudo-code.

\vspace*{-\baselineskip}
\begin{center}
  \begin{minipage}{10cm}
  \begin{tabbing}
    12 \= 12 \= 12 \= 12 \kill \\
    procedure {\sf integrate} {\em level}  \\
    \> do {\sf singlestep} on {\em level}\\
    \> {\sf better boundary} on {\em level}\\
    \> solve {\sf poisson} equation on {\em level} \\ \\
    \> if {\em next level} exists, then \\
    \> \> {\sf default boundary} on {\em next level}\\
    \> \> do r times \\
    \> \> \> {\sf integrate} {\em next level} \\
    \> \> {\sf update} of {\em level} \\
    \> {\sf check} criterion on {\em level}
  \end{tabbing}
\end{minipage}
\end{center}

Starting at time $t_0$, the procedure {\sf singlestep} performs one
timestep $\Delta t_{level}$ on all grids of this level. In addition to
the data within the grid itself the integration scheme needs boundary
data which are by default obtained by interpolation in space and time
from previous level data. In the subsequent procedure {\sf better
  boundary}, the boundary data are, if possible, replaced by values
from neighboring fine grids. After boundary data have been
communicated on each grid, in order to perform the projection step
Poisson equations with fixed boundary are solved for the potentials
$\Delta \psi ^\pm = \omega ^\pm$.  Now all data of the actual level
are advanced to a time $t_0 + \Delta t_{level}$ and the recursion
starts by integrating the next level if existing. The first step in
this recursive process is achieved by supplying information about the
default boundary data from parent grids. This is done by storing the
increments calculated from the actual grids at time $t_0$ and parent
grids at time $t_0 + \Delta t_{level}$. To achieve linear
interpolation in time these increments are added to the boundary data
at the end of {\sf singlestep}. Storing only the increments in a
special C++ boundary class avoids the memory overhead resulting from
keeping data at present and previous times. On this next level, the
timestep $\Delta t_{level+1}$ and the spatial discretization lengths
are divided by a refinement factor~$r$. Therefore, the procedure {\sf
  singlestep} has to be called $r$-times on this new level in order to
reach the time $t_0 + \Delta t_{level}$. Having completed this
recursive integration loop, this level and all finer levels are
advanced to time $t_0 + \Delta t_{level}$. Now the finer level data
are used in procedure {\sf update} to improve the values of the actual
level. The procedure {\sf integrate} is finished by checking if a
certain criterion is fulfilled, that decides whether a refinement step
is performed.

\subsection{Regridding}

The criterion for refinement is adapted to the problem of current
sheet formation. The global maximum of vorticity and current density
is calculated and compared to the values when the last refinement was
done. Regridding is initiated, if the ratio of those maxima exceeds a
prescribed value which is equal to the refinement factor~$r$ due to
the scaling symmetry of the MHD equations~(\ref{mhdz}). The result of
regridding is a new list of levels starting below the actual level.
This new list replaces the old one, which is then deleted.

The logical structure of the {\sf regridding} procedure is shown in
the subsequent pseudo-code.

\vspace*{-\baselineskip}
\begin{center}
  \begin{minipage}{10cm}
    \begin{tabbing}
      12 \= 12 \= 12 \= 12 \kill \\
      procedure {\sf regridding} {\em level} \\
      \>  for all {\em grids} on {\em level} \\
      \> \> mark critical points and append them to a list \\
      \> cover the critical points with rectangles ({\sf saw up}) \\ \\
      \> {\sf nesting} rectangles into their parents and \\
      \> assign parents and neighbors \\ \\
      \> fill the new rectangles with {\sf default data} \\ \\
      \> calculate global maxima for comparison \\
      \> in the procedure {\sf check}\\ \\
      \> if {\em old level} of same resolution existed before \\
      \> regridding, then \\
      \> \> {\sf better data} on {\em new level} from {\em old level} \\
      \> \> solve {\sf poisson} equation on {\em new level} \\
      \> \> if finer {\em level} existed before regridding, then \\
      \> \> \> {\sf regridding} of {\em new level} \\
      \> \> else \\
      \> \> \> assign global maxima from {\em old level} \\
      \> else \\
      \> \> solve {\sf poisson} equation on {\em new level} \\
    \end{tabbing}
  \end{minipage}
\end{center}

The procedure {\sf regridding} starts with a loop over all grids of
{\em level} to collect the critical points. Therefore, we calculate at
each grid point the difference between the convection terms ${\bf
  z}^\mp \cdot \nabla {\bf z}^\pm$ on the actual level and the next
coarser one and if this difference exceeds a prescribed threshold
$\epsilon$, we append this point and a surrounding rectangle of given
size to the list of critical points. In the procedure {\sf saw up}
these critical points are covered with rectangles. The procedure {\sf
  nesting} guarantees that they are properly nested into grids of the
previous level allowing for more than one parent grid. At the same
time, parent and neighbor grids are assigned to each new rectangle.
Since these procedures are the most complex ones, they will be
discussed in detail in the next subsections.

Now as each rectangle has information about his parents, the new
rectangles are filled with spatially interpolated data in {\sf default
  data}. To avoid discontinuities, interpolation is done on the fields
containing the highest derivatives, namely $\nabla \times {\bf
  z}^\pm$.  In order to supply boundary conditions for the solution of
the Poisson equations, data for the potentials $\psi^\pm$ on the
outermost boundary are assigned as well. Afterwards, global maxima
needed in the procedure {\sf check} are calculated.

If the recursive regridding was first invoked on the deepest level,
the procedure is finished by solving the Poisson equations on the new
level. Otherwise, data of the same resolution already existed and are
used in {\sf better data} to get more accurate values for the new
grids. Data for the potentials are available after solving the Poisson
equations. If the old level of same resolution was not the deepest
level, the recursive regridding procedure is applied to the {\em new
  level}. In order to avoid unnecessary rebuilding of the level
hierarchy, global maxima used as reference in {\sf check} are assigned
from the {\em old level} only in the other case.

\subsection{Grid Generation}

The grid generation is performed in the procedure {\sf saw~up} acting
on a list of rectangles. On first entry, this list consists of one
rectangle which covers all critical points of that level. Each
rectangle is now processed in the following way. First, it is decided
in which direction the first cut will take place. Therefore, we
calculate vectors in the $x$- and $y$-direction which contain the
number of critical points in each column or row, respectively.
According to Bell {\it et al.}~\cite{bell-berger-etal:1994}, we call
them horizontal and vertical {\em signatures} $\Sigma$. The first cut
is done in the direction with larger fluctuations in signature. This
is achieved in the procedure {\sf cut dim}, which first seeks for the
best cut in this direction. In the procedure {\sf cut} zeroes of the
signature and its turning points (zeroes of $\Delta_i = \Sigma_{i-1} -
2 \Sigma_i + \Sigma_{i+1}$) are taken into account as possible cuts.
If no such cuts are found, the mid point is chosen. A cut results in
two lists of critical points. Each list is covered by a rectangle of
minimal size.  To every rectangle costs are assigned which are
calculated as a sum of integration and memory costs ($\propto$ the
area), boundary communication costs ($\propto$ the perimeter) and
fixed costs (measuring the overhead for managing one additional grid).
The two rectangles having the minimal costs are returned. Afterwards a
loop over these two rectangles is performed.  They are both given to
the procedure {\sf cut} to find the best cut in the other direction.
The costs of the two new rectangles in comparison to the original
one's are used to decide whether the second cut is accepted or not.
This gives a list of two, three or four rectangles. Their costs are
summed up, and if they are less than the costs of the rectangle which
entered the procedure {\sf cut dim}, they are returned to {\sf saw
  up}.  Otherwise, an empty list is given back.  In the latter case,
if the efficiency measured by the ratio of critical points and grid
points in the rectangle is insufficient, we enforce a cut in the
middle of the longer side of the rectangle. Now the new rectangles are
appended to a temporary list, which is, if not empty, passed to the
recursive procedure {\sf saw up} again. This recursion is stopped when
further cuts do not allow a reduction of costs anymore. The above
treatment is summarized in the two following pieces of pseudo-code.

\vspace*{-\baselineskip}
\begin{center}
  \begin{minipage}{10cm}
    \begin{tabbing}
      12 \= 12 \= 12 \= 12 \kill \\
      procedure {\sf saw up} {\em rectangles} \\
      \> for all {\em rectangles} \\
      \> \> calculate $\Sigma$ and variance in $x$- and $y$-direction \\
      \> \> if variance in $x$ $>$ variance in $y$, then \\
      \> \> \> apply {\sf cut dim} on {\em rectangle} in $x$-direction \\
      \> \> else \\
      \> \> \> apply {\sf cut dim} on {\em rectangle} in $y$-direction \\
      \> \> if no cut found and efficiency insufficient, then \\
      \> \> \> half {\em rectangle} in longer direction \\
      \> \> append resulting {\em rectangles} to temporary list \\
      \> \> {\sf saw up} of temporary list of {\em rectangles} \\
      \> \> if temporary list is not empty, then \\
      \> \> \> replace actual rectangle by temporary list \\
    \end{tabbing}
  \end{minipage}
\end{center}

\vspace*{-2 \baselineskip}
\begin{center}
  \begin{minipage}{10cm}
    \begin{tabbing}
      12 \= 12 \= 12 \= 12 \kill \\
      procedure {\sf cut dim} of {\em rectangle} in direction {\em dim} \\
      \> determine best {\sf cut} in direction {\em dim} \\
      \> and return two rectangles \\ \\
      \> loop over the two rectangles \\
      \> \> cut in other direction \\ \\
      \> \> if costs are smaller than those of actual \\
      \> \> rectangle, then \\
      \> \> \> replace actual rectangle by list \\ \\
      \> compare costs of new list (of 2-4 rectangles) with \\
      \> those of original rectangle and return cheapest
    \end{tabbing}
  \end{minipage}
\end{center}

\begin{figure}[p]
  \begin{center}
    \epsfxsize 6.5cm
    \epsffile{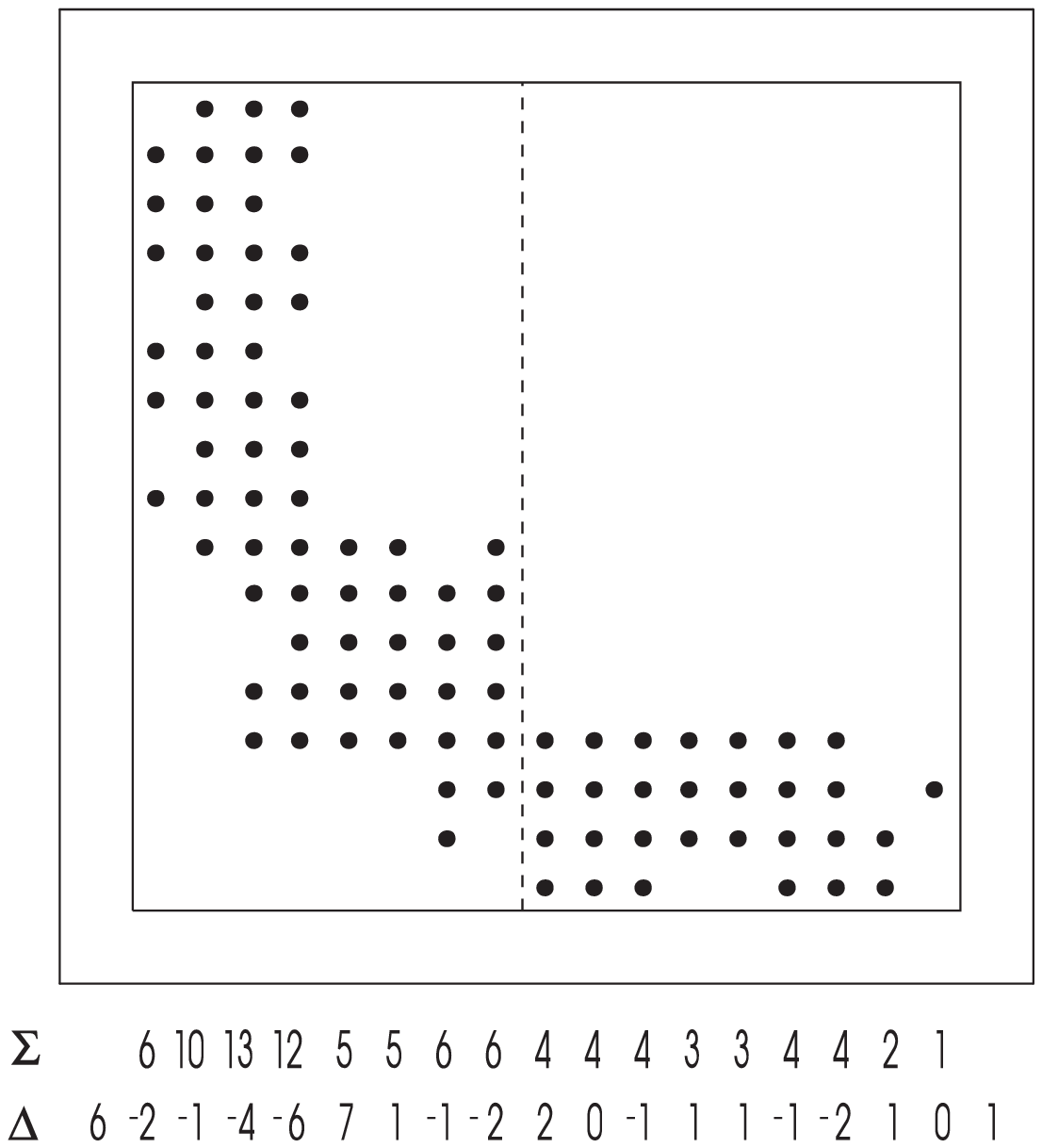}

    \bigskip

    \epsfxsize 6.5cm 
    \epsffile{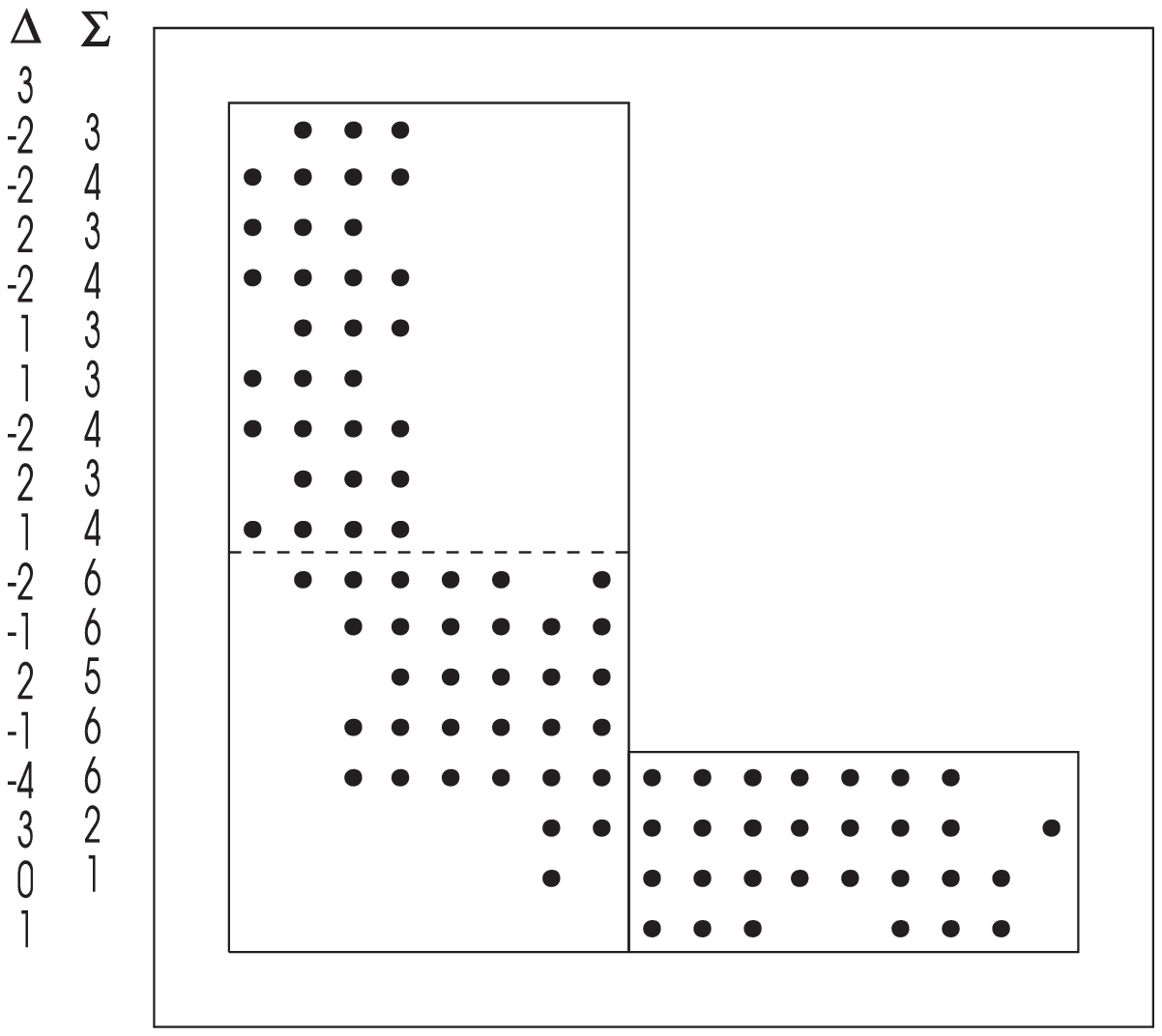}

    \bigskip

    \epsfxsize 6.5cm 
    \epsffile{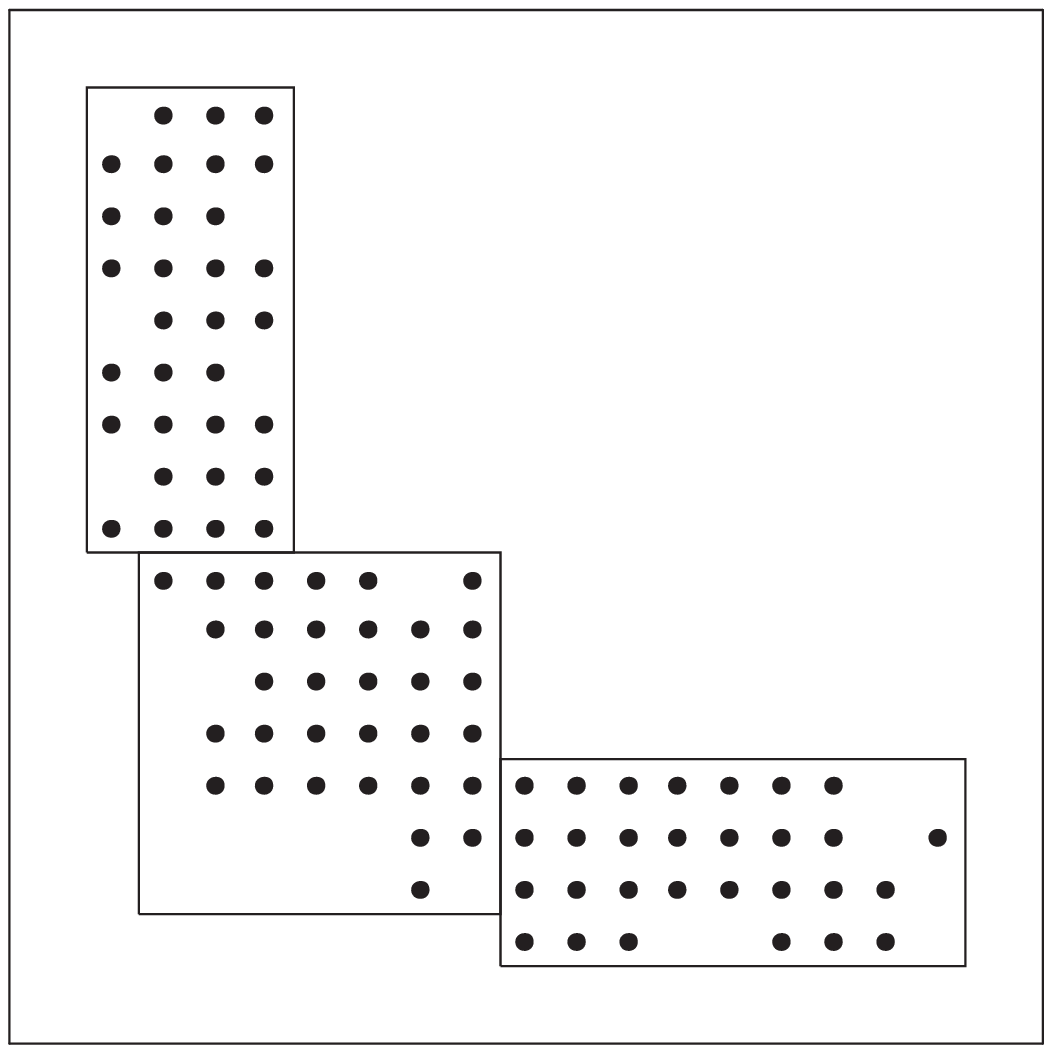}

    \medskip

    \caption{The effect of procedure {\sf saw up}.}
    \label{fig:sawup}

  \end{center}

\end{figure}

An example, where {\sf saw up} produces three new rectangles is shown
in Figure~\ref{fig:sawup}.

\subsection{Nesting}

After generation of non-overlapping rectangles in the procedure {\sf
  saw up}, it is not guaranteed that all rectangles are properly
nested in the rectangles of the parent level. A typical example, where
this is not the case, is shown in Figure~2.

To check, whether a rectangle is properly nested we calculate the sum
of areas of intersections with all rectangles of the parent level.
When this area equals the area of the actual rectangle it is
guaranteed that this rectangle is properly nested. Otherwise, we
proceed as follows. First, we determine the longest common edge of the
just calculated intersections. Our strategy is to avoid coinciding
cuts of several levels. Therefore, we seek for cuts perpendicular to
the longest common edge. Let us assume, as in Figure~2, that this edge
lies in the $y$-direction. Then, we cut where the number of grid
points covered by the intersections in each row changes. This list of
rectangles is recursively tested for proper nesting. Obviously, this
procedure is well suited to assign parents to each rectangle at the
same time. After having obtained a list of properly nested rectangles,
they get information about their neighbors.
\begin{figure}[htbp]
  \begin{center}
    \epsfxsize 6cm
    \epsffile{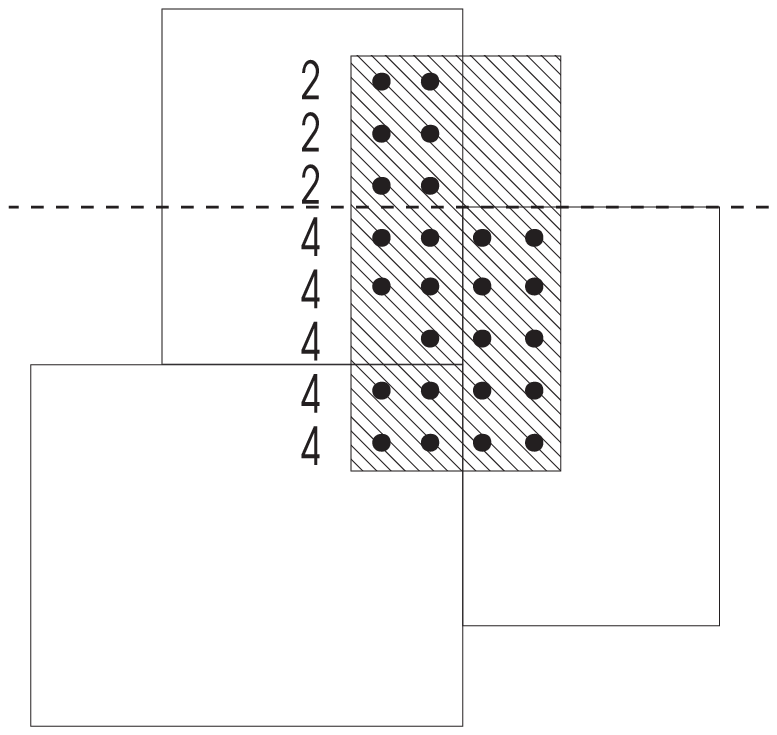}

    \bigskip

    \epsfxsize 6cm
    \epsffile{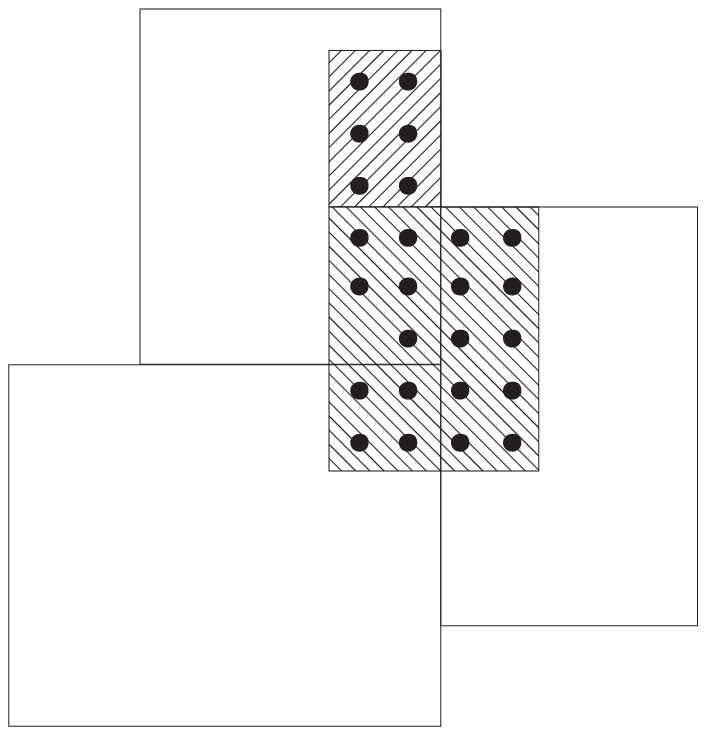}

    \medskip
    \caption{The result of the nesting procedure.}
    \label{fig:nesting}
  \end{center}
\end{figure}

\subsection{Integral and local controls}

In order to extract physical properties of the simulation, it is
necessary to calculate integral quantities like kinetic and magnetic
energy as well as maxima of current density and vorticity. The latter
are easily obtained by looping over all grids and all levels. Integral
quantities are calculated in the following way. First, on the coarsest
level the energy $E_{level}$ (swiss cheese energy) associated to the
area not covered by grids of higher resolution is calculated. This is
repeated down to the lowest level. Finally, the energy is obtained as
a sum over all energies $E_{level}$.

\subsection{Parallelization}

On shared memory machines our adaptive mesh refinement code can be
parallelized in an effective and straight forward way. The main time
of the program is spent in the procedure {\sf singlestep}. Since the
number of grids is much higher than the number of processors
parallelization is done by distributing the grids to the processors.
That means that as soon as a singlestep on a grid is finished, the
next grid is passed to the free processor. This results in a very
effective utilization of all processors. All this can easily be done
using standard Posix threads. The implementation on distributed
memory machines using the shared memory access model is in work.

\section{Numerical Results}

In contrast to simulations of Frisch {\it et
  al.}~\cite{frisch-pouquet-etal:1983} and Sulem {\it et
  al.}~\cite{sulem-frisch-etal:1985}, we choose as initial condition a
modified Orszag--Tang vortex, given by
\begin{eqnarray*}
  \varphi^0(x,y)&=&\cos(x+1.4)+\cos(y+2.0) \ , \\ 
  \psi^0(x,y)&=&\frac{1}{3} \, \left[\cos(2x+2.3)+\cos(y+6.2)\right] \ .
\end{eqnarray*}
This initial condition, which was already used in turbulence
simulations~\cite{biskamp-welter:1989,grauer-marliani:1995}, possesses
less symmetry and is therefore more generic for the formation of
small-scale structures. Computations are done with periodic boundary
conditions on a square of length $2 \pi$. The initial spatial
resolution was given by $256^2$ grid points.

\begin{figure}[tbp]
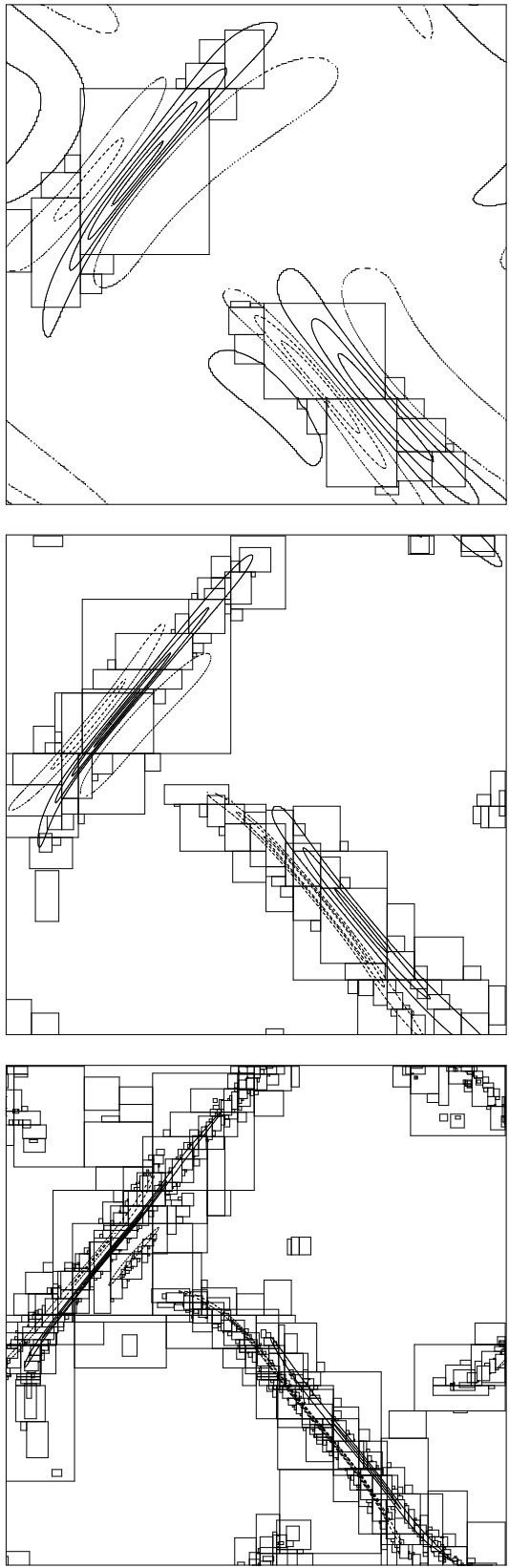

  \begin{center}
    \epsfxsize 7cm
    \rotatebox{-90}{\epsffile{t317.eps}}

    \bigskip
    \epsfxsize 7cm
    \rotatebox{-90}{\epsffile{t430.eps}}

    \bigskip
    \epsfxsize 7cm
    \rotatebox{-90}{\epsffile{t538.eps}}

    \bigskip
    \caption{Evolution of the current density
             at times $1.6$, $2.2$ and $2.7$.}
    \label{fig:contour} 
  \end{center}
\end{figure}

The temporal evolution of the current density is shown in the contour
plots of Figure~\ref{fig:contour}. In addition to the contour levels,
the rectangle hierarchy is plotted. The first plot shows the grid
after the first refinement has taken place. The contour plot at time
$t = 2.2$ contains already 3 levels. At the final time $t = 2.7$ a
total of 5 levels are present. Figure~\ref{fig:contour1} is a contour
plot at the same time as the last one of Figure~\ref{fig:contour}. To
avoid hiding the sharpness of the current sheets no rectangles are
included. In the actual simulation, the refinement factor was equal
to $r=2$. On a workstation with 128 Mbyte of main memory, four
refinements could be realized corresponding to a resolution of
$4096^2$ grid points with a non-adaptive scheme. The limiting factor
is the amount of main memory available, whereas up to this resolution
computational costs are very moderate.

In the first picture the current sheets start to form, afterwards they
evolve into thiner and thiner sheets and the maxima of current density
and vorticity are increasing continually. The current density is
growing exponentially in time. In Figure~\ref{fig:growth} a
semilogarithmic plot of the maximum current density in the upper sheet
is depicted. Included is a fit to an exponential function given by $
j_{\mbox{fit}}(t) = 0.5 \exp(2.115 \, t)$. This functional behavior is
in agreement with the results of Sulem {\it et
  al.}~\cite{sulem-frisch-etal:1985}. A detailed analysis of the
asymptotic scaling behavior and
\twocolumn[{
\begin{figure*}[htbp]
  \widetext
  \begin{center}
    \epsfxsize 12cm
    \rotatebox{-90}{\epsffile{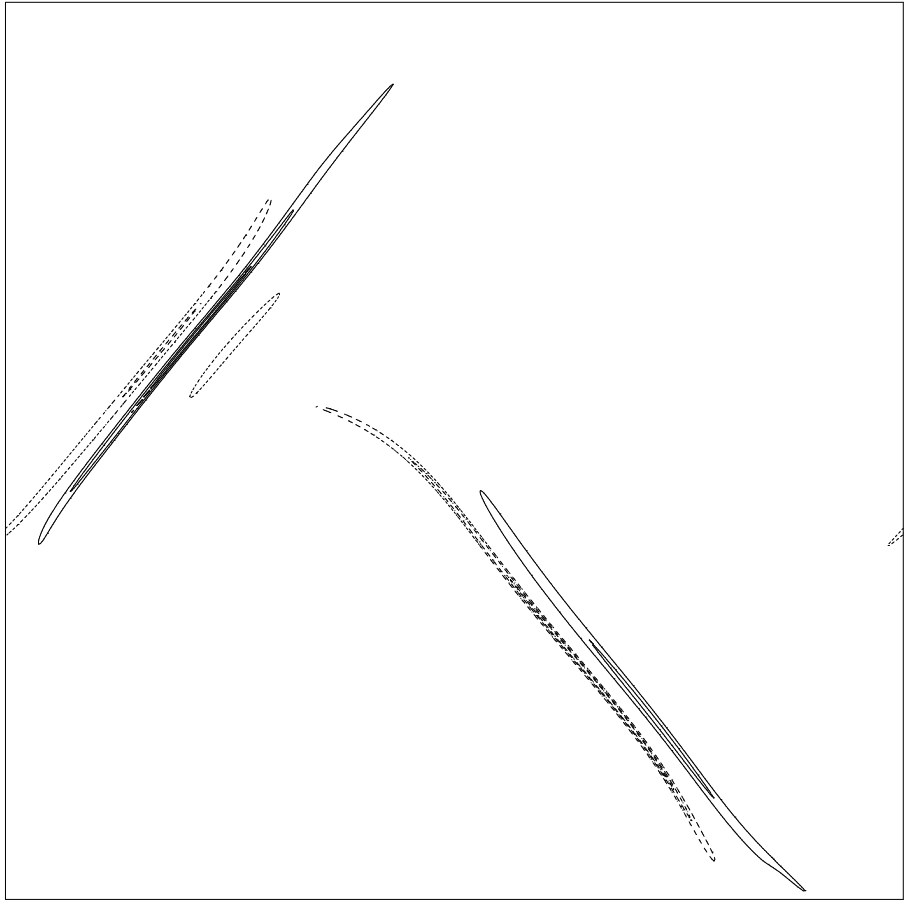}}

    \bigskip
    \caption{Current density at time $2.7$.}
    \label{fig:contour1}
    \narrowtext
  \end{center}
\end{figure*}}]
\noindent a comparison to the
predictions in \cite{sulem-frisch-etal:1985} will be presented
elsewhere.
\begin{figure}[htbp]
  \begin{center}
    \epsfxsize 8cm
    \epsffile{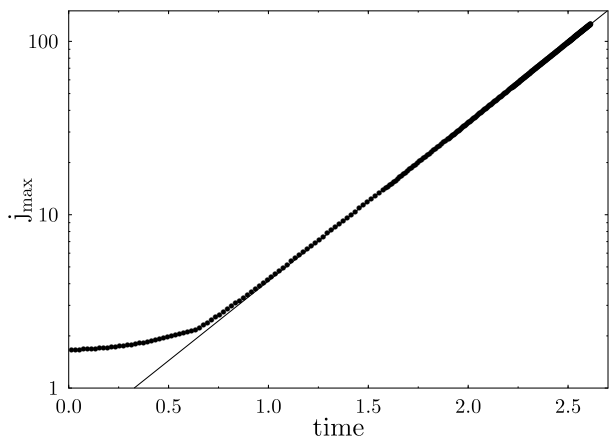}

    \medskip
    \caption{Amplification of current density.}
    \label{fig:growth}
  \end{center}
\end{figure}

\begin{figure}[htbp]
  \begin{center}
    \epsfxsize 8cm
    \epsffile{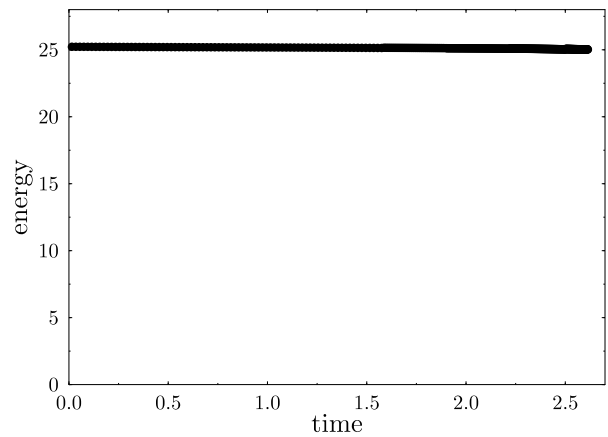}

    \medskip
    \caption{Energy conservation.}
    \label{fig:energy}
  \end{center}
\end{figure}

The second order upwind scheme produces substantial energy dissipation
only if underresolved steep gradients have formed. Therefore, the
energy conservation is a measure whether the singular current sheets
are sufficiently resolved. In Figure~\ref{fig:energy} we give a plot
of energy as a function of time. To be more precise, total energy is
conserved to within less than 1~\%.

\begin{figure}[htbp]
  \begin{center}
    \epsfxsize 8cm
    \epsffile{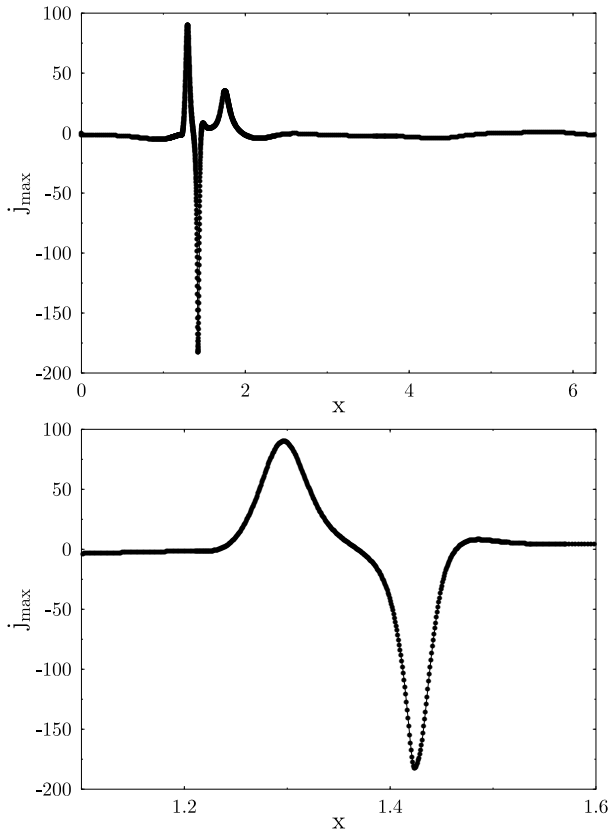}

    \medskip
    \caption{Cuts of current density in $x$-direction through
             the maximum at time $2.72$.}
    \label{fig:section}
  \end{center}
\end{figure}

In order to further illustrate that the current sheets are well
resolved, in Figure~\ref{fig:section} we show one dimensional cuts in
$x$-direction through the maximum of current density in the upper half
of the integration range. In the upper plot the $x$-range equals the
periodicity length. The lower one with a reduced plot range shows that
the grid points of the finest levels very well resolve the current
sheet.

In the previous section we mentioned that a refinement would take
place when the discretization error for the nonlinearity exceeds a
prescribed value $\epsilon$. The choice of the parameter $\epsilon$ is
crucial for the numerical accuracy. If $\epsilon$ is taken too large,
certain regions may be underresolved which can lead to reconnection
and violation of energy conservation. Decreasing systematically the
value of $\epsilon$ has the effect that reconnection phenomena are
suppressed. Below a certain threshold the numerical results proved to
be independent of $\epsilon$. The simulations shown in
Figure~\ref{fig:contour} were performed with $\epsilon = 0.025$.

Applying adaptive mesh refinement to the evolution of singular
structures like current sheets in magnetohydrodynamics is motivated by
the expected reduction of memory needed to resolve them. This is well
justified by the numerical results. To give the reader an impression
of how many grids are generated on the different levels and of the
number of grid points contained in each level's grids, we display
values for the level hierarchy at time $t=2.7$ in the following
tables. In Table~I the results are shown for a simulation with
refinement factor $r=2$ and in II for another one with $r=4$.
\begin{center}
  TABLE I

  Statistics for simulation with $r = 2$. 

  \vspace*{0.3cm}

  \begin{tabular}{rrr} \hline 
    ~level~ & ~~~~~~~number of knots~ & ~~~~~~~grid points in level~ \\ \hline
         0~ &                1~ &                 70225~ \\ 
         1~ &               51~ &                168033~ \\ 
         2~ &              100~ &                341349~ \\ 
         3~ &              178~ &                734426~ \\
         4~ &              417~ &               1557221~ \\ \hline
  \end{tabular}
  \bigskip
\end{center}
\begin{center}
  TABLE II

  Statistics for simulation with $r = 4$. 

  \vspace*{0.3cm}
  
  \begin{tabular}{rrr} \hline
    ~level~ & ~~~~~~~number of knots~ & ~~~~~~~grid points in level~ \\ \hline
         0~ &                1~ &                 70225~ \\ 
         1~ &               49~ &                506073~ \\ 
         2~ &              195~ &               2331952~ \\ \hline
  \end{tabular}
  \bigskip
\end{center}

From level to level the total number of grid points grows much less
than by a factor of $r^2$ necessary for a non-adaptive treatment. For
$r$ chosen equal to $2$, one can see that even for the very small
value of $\epsilon$ prescribed here it increases no more than by a
factor of about $2$. This promises that the compression rate will
improve the more refinements are performed.

In Table~III simulations with different refinement factors are
compared with regard to the total number of grid points on all levels.
The number of grid points on one data field with the same grid spacing
as the finest level in the adaptive code is called the non-adaptive
size. In the last row we give the ratio of the grid points, adaptively
and non-adaptively. For the investigated hierarchy of 5 levels with
refinement factor $r=2$ this ratio is about 17 \%. When the finest
levels are equally resolved, the compression for both refinement
factors is practically indistinguishable. For the comparison of
adaptive versus non-adaptive treatment, the compression rate based on
counting grid points does not fully reflect the total improvement in
main memory consumption. In upwind schemes several auxiliary fields
have to be stored. In non-adaptive simulations these full sized fields
are present all the time whereas here they are needed only temporarily
during the execution of \mbox{\sf singlestep} on a small grid.
\begin{center}
  TABLE III

  Comparison of different refinement factors.

  \vspace*{0.3cm}

  \begin{tabular}{lrr} \hline
                                &  ~~~~~~~~~~$r = 2$ & ~~~~~~~~~~$r = 4$ \\ \hline
    total number of grid points &  2871254 &  2908250 \\
    non-adaptive size           & 16851025 & 16851025 \\
    ratio                       &    0.170 &    0.172 \\ \hline
  \end{tabular}
  \bigskip
\end{center}                                

\begin{figure}[htbp]
  \begin{center}
    \epsfxsize 8cm \epsffile{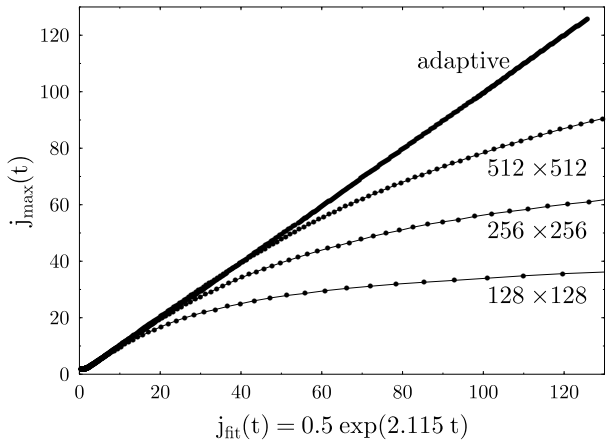}

    \medskip
    \caption{Comparison of adaptive and non-adaptive simulations.}
    \label{fig:compare}
  \end{center}
\end{figure}

We want to finish this section with an impressive comparison of the
results for the amplification of the current density for several
non-adaptive grid sizes and for the adaptive code.
Figure~\ref{fig:compare} is a parametric plot of the maximum of
current density as a function of the fit $j_{\mbox{\scriptsize
    fit}}(t) = 0.5 \exp(2.115 \, t)$ already depicted in
Figure~\ref{fig:growth}. In addition to the results of the adaptive
mesh refinement code we include data obtained with fixed grids of
resolutions $128^2$, $256^2$ and $512^2$. Until the simulations become
underresolved, a linear behavior is also observed in the non-adaptive
simulations. Then the upwind method introduces numerical viscosity
leading to reconnection processes and substantial energy dissipation.

\section{Conclusions}

The complexity of adaptive mesh refinement compared to non-adaptive
treatments should not be underestimated. On the other hand, the
growing progress of object oriented programming languages helps
enormously to reduce the difficulties in programming. To give some
impression, the programs needed for regridding, nesting and the
handling of data structures are only about 3000 lines of C++ code.

As we have demonstrated, adaptive mesh refinement is a powerful tool
to study the evolution of singular structures as the formation of
current sheets in ideal MHD. Other problems of this type like in the
axisymmetric \cite{grauer-sideris:1995} and the full three dimensional
Euler equations are natural candidates for this method. Work in this
direction is in progress.

Whether adaptive mesh refinement is also a useful concept for
simulating turbulent hydro- and magnetohydrodynamic flows will depend
on how efficiently the small scale structures can be covered by
hierarchically nested grids.

\section*{Acknowledgment}
We like to thank K. H. Spatschek for his continuous support.  This
work was performed under the auspices of the Sonderforschungsbereich
191.

\bibliographystyle{phjcp}

\begin{thebibliography}{10}

\bibitem{beale-kato-etal:1984}
{\sc J.~T. Beale}, {\sc T.~Kato}, and {\sc A.~Majda},
\newblock {\em Comm. Math. Phys.} {\bf 94}, 61 (1984).

\bibitem{bell-marcus:1992b}
{\sc J.~B. Bell} and {\sc D.~L. Marcus},
\newblock {\em Comm. Math. Phys.} {\bf 147}, 371 (1992).

\bibitem{kerr:1993}
{\sc R.~M. Kerr},
\newblock {\em Phys. Fluids A} {\bf 5}, 1725 (1993).

\bibitem{pumir-siggia:1990}
{\sc A.~Pumir} and {\sc E.~Siggia},
\newblock {\em Phys. Fluids. A} {\bf 2}, 220 (1990).

\bibitem{pumir-siggia:1992}
{\sc A.~Pumir} and {\sc E.~Siggia},
\newblock {\em Phys. Fluids. A} {\bf 4}, 1472 (1992).

\bibitem{berger-colella:1989}
{\sc M.~J. Berger} and {\sc P.~Colella},
\newblock {\em J. Comput. Phys.} {\bf 82}, 64 (1989).

\bibitem{bell-berger-etal:1994}
{\sc J.~Bell}, {\sc M.~Berger}, {\sc J.~Saltzman}, and {\sc M.~Welcome},
\newblock {\em SIAM J. Sci. Comput.} {\bf 15}, 127 (1994).

\bibitem{bell-colella-etal:1989}
{\sc J.~B. Bell}, {\sc P.~Colella}, and {\sc H.~M. Glaz},
\newblock {\em J. Comput. Phys.} {\bf 85}, 257 (1989).

\bibitem{bell-marcus:1992}
{\sc J.~B. Bell} and {\sc D.~L. Marcus},
\newblock {\em J. Comput. Phys.} {\bf 101}, 334 (1992).

\bibitem{grauer-marliani:1995}
{\sc R.~Grauer} and {\sc C.~Marliani},
\newblock {\em Phys. Plasmas} {\bf 2}, 41 (1995).

\bibitem{sulem-frisch-etal:1985}
{\sc P.~L. Sulem}, {\sc U.~Frisch}, {\sc A.~Pouquet}, and {\sc M.~Meneguzzi},
\newblock {\em J. Plasma Phys.} {\bf 33}, 191 (1985).

\bibitem{frisch-pouquet-etal:1983}
{\sc U.~Frisch}, {\sc A.~Pouquet}, {\sc P.~L. Sulem}, and {\sc M.~Meneguzzi},
\newblock {\em J. M\'{e}c. Th\'{e}or. Appl.} , 191 (1983).

\bibitem{biskamp-welter:1989}
{\sc D.~Biskamp} and {\sc H.~Welter},
\newblock {\em Phys. Fluids B} {\bf 1}, 1964 (1989).

\bibitem{grauer-sideris:1995}
{\sc R.~Grauer} and {\sc T.~C. Sideris},
\newblock {\em Physica~D} {\bf 88}, 116 (1995).

\end{thebibliography}

\end{document}